# A novel $\sqrt{19} \times \sqrt{19}$ superstructure in epitaxially grown 1*T*-$TaTe_2$

*Jinwoong Hwang*†*, *Yeongrok Jin*†, *Canxun Zhang*†, *Tiancong Zhu*†, *Kyoo Kim, Yong Zhong, Ji-Eun Lee, Zongqi Shen, Yi Chen, Wei Ruan, Hyejin Ryu, Choongyu Hwang, Jaekwang Lee, Michael F. Crommie, Sung-Kwan Mo* and Zhi-Xun Shen**

Dr. J. Hwang, Dr. Y. Zhong, Prof. Z.-X. Shen
Stanford Institute for Materials and Energy Sciences, SLAC National Accelerator Laboratory, Menlo Park, CA, USA
E-mail: jinwoonghwang@lbl.gov, zxshen@stanford.edu

Prof. Z.-X. Shen
Geballe Laboratory for Advanced Materials, Department of Physics and Applied Physics, Stanford University, Menlo Park, CA, USA
E-mail: zxshen@stanford.edu

Dr. J. Hwang, Dr. Y. Zhong, J. Lee, Dr. S.-K, Mo
Advanced Light Source, Lawrence Berkeley National Laboratory, Berkeley, CA, USA
E-mail: jinwoonghwang@lbl.gov, skmo@lbl.gov

C. Zhang, Dr. T. Zhu, Z. Shen, Dr. Y. Chen, Prof. W. Ruan, Prof. M. F. Crommie
Department of Physics, University of California, Berkeley, CA, USA

C. Zhang, Dr. T. Zhu, Dr. Y. Chen, Prof. W. Ruan, Prof. M. F. Crommie
Materials Sciences Division, Lawrence Berkeley National Laboratory, Berkeley, CA, USA

C. Zhang, Prof. M. F. Crommie
Kavli Energy NanoScience Institute, University of California, Berkeley, CA, USA

J. Lee, Dr. H. Ryu
Center for Spintronics, Korea Institute of Science and Technology, Seoul, South Korea

Prof. W. Ruan
Department of Physics, Fudan University, Fudan, China

Dr. J. Hwang, Y. Jin, J. Lee, Prof. C. Hwang, Prof. J. Lee
Department of Physics, Pusan National University, Busan, South Korea
E-mail: jinwoonghwang@lbl.gov

Prof. C. Hwang
Quantum Matter Core-Facility, Pusan National University, Busan, South Korea

Dr. K. Kim
Korea Atomic Energy Research Institute, Daejeon, South Korea

**The spontaneous formation of electronic orders is a crucial element for understanding complex quantum states and engineering heterostructures in two-dimensional materials. We report a novel $\sqrt{19} \times \sqrt{19}$ charge order in few-layer thick 1*T*-$TaTe_2$ transition metal dichalcogenide films grown by molecular beam epitaxy, which has not been realized. Our photoemission and scanning probe measurements demonstrate that monolayer 1*T*-$TaTe_2$ exhibits a variety of metastable charge density wave orders, including the $\sqrt{19} \times \sqrt{19}$ superstructure, which can be selectively stabilized by controlling the post-growth annealing temperature. Moreover, we find that only the $\sqrt{19} \times \sqrt{19}$ order persists in 1*T*-$TaTe_2$ films thicker than a monolayer, up to 8 layers. Our findings identify the previously unrealized novel electronic order in a much-studied transition metal dichalcogenide and provide a viable route to control it within the epitaxial growth process.**

The enhanced electron-electron and electron-phonon interactions due to quantum confinement and reduced screening in two-dimensional (2D) materials often result in complex electronic phases distinct from the corresponding bulk systems [1,2]. One of the most prominent examples is the spontaneous breaking of translational symmetry by forming a charge density wave (CDW) order [1,3]. Transition metal dichalcogenides (TMDCs) are an ideal testbed to investigate the various types of the CDW order and their origins [1-3], since different CDW orders can emerge from a limited number of structural polytypes, and often competes with other orders such as superconductivity and magnetism [2-5]. Among TMDCs, $TaX_2$ (X = S, Se) has been extensively studied as a prototypical material to explore the role of strong electron correlation in an electron-phonon coupling driven CDW system [6-11]. For example, 2*H*-$TaX_2$ (trigonal-prismatic coordination) shows $3 \times 3$ CDW order (**Figure 1a**) [1,10,11], where the outer six Ta atoms move towards the center Ta atom forming a seven-atom-cluster, while the other

Ta atoms remain at the same positions. $\sqrt{13} \times \sqrt{13}$ CDW order (**Figure 1b**) [1,8-10] is the more stable phase for 1*T*-$TaX_2$ (octahedral coordination), in which the Ta atoms reposition themselves to form a Star-of-David pattern. The CDW orders in $TaX_2$ persist down to the monolayer (ML) limit [8-11]. At the same time, exotic quantum phases emerge due to the strong electron-electron interaction, including unusual orbital textures [8] and quantum spin liquids behavior [9] in Mott insulating ML 1*T*-$TaSe_2$.

Despite being in the same family of Ta-based dichalcogenides, 1*T*-$TaTe_2$ has been relatively less investigated. In this material, the stronger Te-Te interlayer coupling would likely create distinct electronic and structural features compared to those of $TaS_2$ and $TaSe_2$ [12-16]. Indeed, charge orders that are rarely seen in other TMDCs, such as $3 \times 1 \times 3$ double zigzag chain (**Figure 1c**) and $3 \times 3 \times 3$ strings of a butterfly-like cluster (**Figure 1d**) [15-19], have been observed in bulk 1*T*-$TaTe_2$, in contrast to the $3 \times 3$ and $\sqrt{13} \times \sqrt{13}$ CDW orders (**Figures 1a** and **1b**) of $TaS_2$ and $TaSe_2$. A natural question arises regarding the type of CDW order that may emerge when the strong Te-Te interlayer coupling gets completely removed in ML 1*T*-$TaTe_2$. It may exhibit a charge order similar to those of ML $TaS_2$ and $TaSe_2$, as is the case for ML 1*T*-$TiTe_2$ [20,21]. It is also possible that the electronic structure of ML 1*T*-$TaTe_2$ is completely modified in the absence of the Te-Te interlayer coupling and results in a distinct charge-ordered state, as is the case for 1*T*-$IrTe_2$. A large band gap with a unique $2 \times 1$ dimer structure was found in ML 1*T*-$IrTe_2$, in stark contrast to the metallic multilayers and bulk [22].

Here, we report a successful growth of 1*T*-$TaTe_2$ thin films using molecular beam epitaxy (MBE) on bilayer graphene (BLG)-terminated 6*H*-SiC(0001) substrate. We have characterized its atomic and electronic structures by reflection high-energy electron diffraction (RHEED), angle-resolved photoemission spectroscopy (ARPES), and scanning tunneling microscopy/spectroscopy (STM/STS). Our experimental results reveal that ML 1*T*-$TaTe_2$ shows a variety of metastable CDW orders, including $3 \times 3$, $\sqrt{13} \times \sqrt{13}$, and $\sqrt{19} \times \sqrt{19}$ superstructures, which can be selectively stabilized by controlling the post-growth annealing temperatures. In particular, the $\sqrt{19} \times \sqrt{19}$ CDW order (**Figure 1e**) has not been realized in any TMDC materials. We find that multilayer 1*T*-$TaTe_2$ up to 8 layers shows only $\sqrt{19} \times \sqrt{19}$ CDW order, distinct from ML and bulk form of 1*T*-$TaTe_2$. Our findings establish the epitaxially grown few-layer 1*T*-$TaTe_2$ thin films as a unique platform to construct and investigate a novel $\sqrt{19} \times \sqrt{19}$ CDW order in layered 2D materials.

**Figure 2a** shows a RHEED image taken at room temperature (RT) for a sub-ML coverage of 1*T*-$TaTe_2$ on BLG substrate grown at lower than 280 ˚C. Sharp vertical lines and the oscillation in their intensity represent a layer-by-layer growth mode. Using the lattice constant of BLG (2.46 Å) as a reference, we can estimate the lattice constant of ML 1*T*-$TaTe_2$ on BLG substrate to be 3.70 ± 0.02 Å. A large-scale STM topographic image in **Figure 2d** shows the typical morphology of a high-quality epitaxially grown ML 1*T*-$TaTe_2$ film. Interestingly, when the ML 1*T*-$TaTe_2$ film is annealed over 340 ˚C after the growth, two additional faint vertical lines (green arrows in **Figure 2b**) show up between two sharp main lines, which are absent in **Figure 2a**. This type of additional faint lines in the RHEED is generally originated from superstructures resulting from surface reconstructions [23,24] or CDWs [25]. Further increasing the annealing temperature over 380 ˚C, even more additional lines (four cyan arrows in **Figure 2c**) show up between the main lines. The appearance of the additional RHEED lines is an irreversible process. Once we obtain the more complex pattern by annealing at higher temperature, it does not turn back to the simpler pattern by annealing at lower temperature again (**Figure S1**). **Figures 2e** and **2f** represent core level PES spectra for Ta 4*f* and Te 4*d*, respectively. The core level measurements clearly show peak shifts depending on the annealing temperatures. The Ta 4*f* (Te 4*d*) peaks move towards lower (higher) binding energy with increasing annealing temperature, indicating the changes in the local atomic coordinate of both Ta and Te.

To further investigate the nature of the changes observed in the RHEED and the core level PES of ML 1*T*-$TaTe_2$, we performed *in situ* polarization-dependent ARPES measurements. **Figures 3a-3h** present ARPES intensity maps of ML 1*T*-$TaTe_2$ along the M−Γ−M direction with different annealing temperatures. The spectra were taken using *s*- (**Figures 3a-3d**) and *p*-polarized photons (**Figures 3e-3h**) at 13 K. The low energy electronic structure near Fermi energy ($E_F$) for $T_{anneal}$ = 265 ˚C sample is best captured in the ARPES intensity maps measured with *p*-polarized photons (**Figure 3e**). There are two bands crossing $E_F$ near the Γ point, and the intensity of the outer band is much stronger than that of the inner band. Changing the photon polarization to in-plane (*s*-polarized photons), the PES signal from the two bands crossing $E_F$ observed using *p*-polarized photons is mostly diffused (**Figure 3a**). Instead, two hole bands become more pronounced, one crossing $E_F$ with a weak intensity (red arrow) and the other with a band maximum at ~0.2 eV below $E_F$ (black arrow). With increasing annealing temperature, the intensity of the hole band crossing $E_F$ (red arrow in **Figure 3a**) becomes much stronger. In contrast, the intensity of the other hole band (black arrow in **Figure 3a**) weakens and becomes almost completely suppressed when reaching $T_{anneal}$ = 340 ˚C (**Figures 3b,3c,3f, and 3g**). The experimental band structure for $T_{anneal}$ = 340 ˚C matches well with the calculated electronic structure within the density functional theory (DFT), with coexisting $3 \times 3$ and $\sqrt{13} \times \sqrt{13}$ CDW orders (**Figure S2**). This result is consistent with the RHEED pattern as shown in **Figure 2b**, in which two additional faint vertical lines emerge only after annealing to $T_{anneal}$ = 340 ˚C. However, the ARPES data for $T_{anneal}$ = 265 ˚C (**Figures 3a** and **3e**) do not show good correspondence with any of the DFT calculations made for undistorted 1*T* structure, 3 × 3, or $\sqrt{13} \times \sqrt{13}$ CDW orders, instead STM topograph shows incommensurate 2.7 × 2.7 superstructure with strong disorder for $T_{anneal}$ = 265 ˚C (**Figure S3**).

When the 1*T*-$TaTe_2$ film is annealed over 380 ˚C, the ARPES band structures show further significant modifications. A new concave band that extends to higher momentum appears at the energy ~0.6 eV below $E_F$ (cyan arrow in **Figure 3h**), while the hole band with the maximum around ~ 0.2 eV and the flat band at 1eV in the case of $T_{anneal}$ = 340 ˚C (white arrows in **Figure 3g**) disappear entirely. According to our DFT calculations, both features that disappear after high-temperature annealing are related to the 3 × 3 CDW order (**Figure S2**). For the *s*-polarized ARPES intensity map (**Figure 3d**), the overall intensity is dominated by the hole band near $E_F$ (**Figure S4**), while most of the other band features fade away, especially at energies higher than 1 eV. The changes in the ARPES band structures with high-temperature annealing at $T_{anneal}$ = 380 ˚C correspond to the emergence of the additional superstructure peaks in the RHEED pattern (**Figure 2c**). A careful investigation of the BLG $\pi$ band reveals that the influence from BLG substrate such as charge transfer and strain is negligible to the overlaid ML 1*T*-$TaTe_2$ film (**Figures S5 and S6**).

To uncover the nature of the possible new superstructure emerging from high-temperature annealing, we performed STM measurements, which can provide direct real-space information of CDW orders to complement the RHEED and ARPES results. An atomically resolved STM topograph for ML 1*T*-$TaTe_2$ film annealed at 380 ˚C clearly shows a $\sqrt{19} \times \sqrt{19}$ crystalline structure (**Figure 3i**), in which six Ta atoms move towards the center Ta atom to form a seven-atom-cluster surrounded by six triangular clusters of Ta (**Figure 1e**). The Fourier transform (FT) of the STM topograph (**Figure 3j**) confirms the commensurate $\sqrt{19} \times \sqrt{19}$ crystalline structure, and further reveals the lattice parameters of the atomic unit cell $a_{\mathrm{lattice}}$ = 3.70 Å, and the CDW unit cell, $a_{\mathrm{CDW}}$ = 16.1 Å (with the convention of $\theta$ = 120° primitive cell). Moreover, STS measurements reveal significant change of the d$I$/d$V$ signal depending on the CDW orders (**Figure S7**). The ARPES band structures from the $T_{\mathrm{anneal}}$ = 380 °C samples (**Figures 3d** and **3h**) exhibit good correspondence to the unfolded band calculations based on the identified $\sqrt{19} \times \sqrt{19}$ crystal structure from the STM measurement (**Figure S8**), thus confirming the emergence of a $\sqrt{19} \times \sqrt{19}$ CDW order in ML 1*T*-$TaTe_2$.

Some of the novel CDW orders in TMDCs only emerge in the ML limit owing to the quantum confinement [22,26-29]. We have grown multilayer 1*T*-$TaTe_2$ films up to 8 layers (L) to examine whether the $\sqrt{19} \times \sqrt{19}$ CDW order is only limited to the ML 1*T*-$TaTe_2$. **Figures 4a-4h** show ARPES data of multilayer 1*T*-$TaTe_2$ films, taken along the M–Γ–M direction using *s*- (**Figures 4a-4d**) and *p*-polarized photons (**Figures 4e-4h**). The overall band structure remains essentially the same (**Figures 4a-4h**), except for some additional features due to the quantum confinement effect, such as flat bands and band splitting near $E_F$ coming from interlayer coupling [20,21,30,31]. The atomically resolved STM topograph for the two-layer 1*T*-$TaTe_2$ film still shows the $\sqrt{19} \times \sqrt{19}$ superstructure (**Figure S9**), and the thickness-dependent core level measurements for Ta 4*f* and Te 4*d* (**Figures 4i and 4j**) do not show any shifts from 2 L to 8 L samples, while the peak positions and the shapes are distinct from those of the bulk. All of these findings indicate that the $\sqrt{19} \times \sqrt{19}$ CDW order is robust in epitaxially grown multilayer 1*T*-$TaTe_2$ films up to 8 L. No evidence of strain by substrate and its relaxation with increasing thickness of the film has been observed (**Figure S6**). In addition, we have found that the two-layer 1*T*-$TaTe_2$ film does not show any changes in the ARPES and the core level spectra regardless of the annealing temperature (**Figure S10**), in contrast to the ML case, suggesting that the $\sqrt{19} \times \sqrt{19}$ CDW order is the most stable order in epitaxially grown multilayers films of 1*T*-$TaTe_2$. Instead, $T_{anneal}$ only affects the quality of the epitaxial films in multilayer 1*T*-$TaTe_2$, rather than affecting the microstructures, for example, through the change in the Te-Te interlayer coupling (**Figures S10 and S11**). Moreover, we do not find any sign of nearly commensurate or incommensurate CDW order, which is well known for the bulk 1*T*-$TaS_2$ [32,33], and only commensurate $\sqrt{19} \times \sqrt{19}$ CDW order is obtained in 1*T*-$TaTe_2$ up to 300 K (**Figure S12**).

**Discussion**

To further understand the experimentally obtained multiple CDW ground states of ML 1*T*-$TaTe_2$, we performed first-principles calculations for the phonon dispersions and total energy differences for all possible CDW superstructures. The phonon calculations found that 3 × 3, $\sqrt{13} \times \sqrt{13}$, and $\sqrt{19} \times \sqrt{19}$ CDW orders are all stable in ML 1*T*-$TaTe_2$ (**Figure S13**). However, we found a clear sign of phonon instability in the natural hypothetical high-temperature unit cell of undistorted 1*T*-$TaTe_2$ (**Figure S14i**), suggesting its susceptibility to form a CDW order. The total energy differences and unit cell volumes for 3 × 3, $\sqrt{13} \times \sqrt{13}$, and $\sqrt{19} \times \sqrt{19}$ CDW orders are given in Table I. The total energy difference of CDW states is defined as the difference (per formula unit) between the energy of the relaxed CDW supercell and the energy of the undistorted ML 1*T* structure. Although 3 × 3 CDW order exhibits the most stable state in ML 1*T*-$TaTe_2$, the others also have quite large energy gains (-69 ~ -106 meV) compared to those of 1*T*-$TaS_2$ and 1*T*-$TaSe_2$ (-13 ~ -67 meV). These results explain why ML 1*T*-$TaS_2$ and 1*T*-$TaSe_2$ only show $\sqrt{13} \times \sqrt{13}$ CDW order, namely the instability of 3 × 3 structure in 1*T*-$TaS_2$ and the large relative differences in energy gains favoring $\sqrt{13} \times \sqrt{13}$ CDW order in both 1*T*-$TaS_2$ and 1*T*-$TaSe_2$.

The total energy differences for $3 \times 3$ and $\sqrt{13} \times \sqrt{13}$ structures in ML 1*T*-$TaTe_2$ are very close, while $\sqrt{19} \times \sqrt{19}$ CDW order shows a smaller energy gain. However, the energy lowered by forming the $\sqrt{19} \times \sqrt{19}$ structure is still much large in its absolute value compared to those of ML 1*T*-$TaS_2$ and 1*T*-$TaSe_2$. The total energy differences summarized in Table I are consistent with our experimental observations in RHEED, ARPES, and STM, where $3 \times 3$ and $\sqrt{13} \times \sqrt{13}$ CDW orders are simultaneously formed within the conventional low-temperature annealing process since they are the most stable with a small difference in the total energy gain. The $\sqrt{19} \times \sqrt{19}$ CDW order is only possible when enough thermal energy is provided, through a high-temperature annealing process, to overcome the hill in the free energy space and stabilize the system in a new local valley.

Despite the stable local minimum energy, the emergence of $\sqrt{19} \times \sqrt{19}$ CDW order in 1*T*-$TaTe_2$ is quite surprising and unexpected [34,35], as it has been rarely seen in any other TMDC. Our DFT calculations, including phonon dispersion, electronic susceptibility, and electron-phonon coupling for the undistorted ML 1*T*-$TaTe_2$ (**Figure S14**), do not find any conventional connection to $\sqrt{19} \times \sqrt{19}$ CDW transition, such as Peierls instability or momentum dependent strong electron-phonon coupling [1,2]. Moreover, the spin-orbit coupling (SOC) does not dramatically alter the low energy electronic structure of the undistorted ML 1*T*-$TaTe_2$ and preserves the overall band structure near $E_F$ and Fermi surface topology despite the expected strong SOC from heavy Ta and Te atoms [34]. What stands out in the electronic structure and phonon calculations of undistorted ML 1*T*-$TaTe_2$, against ML 1*T*-$TaS_2$ and 1*T*-$TaSe_2$, are the sharp peak in the density of states (DOS) right at $E_F$ and additional hole pocket in the Fermi surface (**Figure S14**). As the chalcogen size increases, the energy of the chalcogen *p* states moves closer to the lower Ta *d* states [12-14], which enhances the hybridization between these states. The large DOS at $E_F$ in ML 1*T*-$TaTe_2$ is due to such overlap and mixing of *p*-*d* orbitals, and it can trigger the Jahn-Teller instability to reduce the total energy of the system [36-39], resulting in multiple commensurate CDW states including $\sqrt{19} \times \sqrt{19}$ structure. The diversity and large total energy gain of 1*T*-$TaTe_2$ CDW states are quite remarkable and may be attributed to the stronger tendency towards the change in bonding angles due to the larger polarizability (or low electronegativity) of Te [12,13,34-37].

Distinct CDW orders in ML have been observed in many TMDCs [20,22,26-29], but adding another layer often immediately suppresses them by recovering the interlayer coupling and restoring the symmetry [20,22]. However, we found that not only our epitaxially grown 1*T*-$TaTe_2$ multilayers show the robustness of $\sqrt{19} \times \sqrt{19}$ CDW order up to 8 L (**Figure 4**), but the results are distinct from $3 \times 1 \times 3$ double zigzag chains (**Figure 1c**) and $3 \times 3 \times 3$ strings of butterfly-like cluster (**Figure 1d**) periodicities of the bulk [12-19]. We also note that ultrathin films of 1*T*-$TaTe_2$ grown by chemical vapor deposition (CVD) have been identified to have 1*T*' structure [40].

There are a few interesting points worth mentioning to account for such differences. First is the essential difference in the growth process between the epitaxial layer-by-layer growth and bulk single-crystal growth. To form the trimerization with peculiar chemical $\sigma$-bonding among $t_{2g}$ *d*-orbitals connecting three adjacent Ta sites in bulk $3 \times 1 \times 3$ structure, 1/3 electron needs to be transferred to the Ta site by forming a strong Te-Te interlayer coupling [12-14]. However, the 1*T*-$TaTe_2$ layer in the layer-by-layer growth mode already has clustered Ta even in the ML due to the unstable ML 1*T*-$TaTe_2$ (**Figures S1** and **S14**). Restored Te-Te interlayer coupling in epitaxial multilayer growth may not be strong enough to lift up the Ta clustering and transfer extra electrons to Ta. On the other hand, in the bulk growth mode, single crystals form in the undistorted 1*T*-$TaTe_2$ phase at a much higher temperature [19] (over 500 ˚C ; higher than 3 × 1 × 3 structural transition temperature [41,42]) without pre-clustered Ta. The second is partial intercalation of Ta atoms, which is often found in multilayer TMDC films [43-46]. For example, vanadium (V)-based dichalcogenides show 1*T*' structure in the epitaxially grown multilayer films induced by the intercalated V atoms [44-46]. The 1*T*' structure in CVD-grown ultrathin 1*T*-$TaTe_2$ films [40] may originate from the intercalation of Ta atoms. In our case, the shoulder peaks in the Te 4*d* core levels (blue arrows in **Figure 4j**) for 8 L film may indicate some amount of Ta being intercalated. However, since the main peak position stays the same, and ARPES and RHEED still show $\sqrt{19} \times \sqrt{19}$ CDW order. The difference may be related to the amount of the intercalated Ta atoms [43] that is supposed to be much smaller in the MBE growth. Third, the thickness to realize the crossover from the thin film to bulk has not been met in our study even at 8 L, thicker than most of the other TMDC ultrathin films [30,31,47]. These points need to be addressed by further experiments such as the partial intercalation to bulk 1*T*-$TaTe_2$ crystals and the optimization of growth conditions to achieve thicker films by MBE.

In summary, we have successfully synthesized a few atomic layer thick 1*T*-$TaTe_2$ thin films on a BLG substrate by MBE. Our combined RHEED, ARPES, STM, and first-principles calculations study has revealed multiple CDW orders in ML 1*T*-$TaTe_2$, including a previously unrealized $\sqrt{19} \times \sqrt{19}$ CDW order. We have also shown that controlling the post-growth annealing temperature can selectively stabilize one of the CDW orders. Moreover, the thickness-dependent study finds the robustness of $\sqrt{19} \times \sqrt{19}$ CDW order in multilayer films up to 8 layers. Our findings suggest that the layer-by-layer growth of TMDC films by MBE can be a unique tool to create novel electronic orders in 2D materials that are hard to attain otherwise.

**Experimental Section**

*Thin film growth and in situ ARPES measurement*: 1*T*-$TaTe_2$ thin films were grown by MBE on epitaxial bilayer graphene on 6*H*-SiC(0001). They were then transferred without leaving an ultra-high vacuum (UHV) environment into the ARPES analysis chamber for the measurement at the HERS endstation of Beamline 10.0.1, Advanced Light Source, Lawrence Berkeley National Laboratory. The base pressure of the MBE chamber was $3 \times 10^{-10}$ Torr. High-purity Ta (99.9%) and Te (99.999%) were evaporated from an e-beam evaporator and a standard Knudsen effusion cell, respectively. The flux ratio was Ta:Te = 1:30, and the substrate temperature was held at 265 ˚C for monolayer and 380 ˚C for multilayers 1*T*-$TaTe_2$ during the growth. This yields the growth rate of 45 minutes per monolayer monitored by *in situ* RHEED, and only 8 L film was grown with the growth rate of 15 minutes per monolayer with the flux ratio Ta:Te = 1:20. We carried out *in situ* RHEED measurements with a high voltage of 20 kV throughout the growth process. ARPES data were taken using a Scienta R4000 analyzer at a base pressure $2 \times 10^{-11}$ Torr. The photon energy was set at 55 eV for *s*-polarization and 83 eV for *p*-polarization with energy and angular resolution of 15-25 meV and 0.1˚, respectively. The spot size of the photon beam on the sample was ~100 μm x 100 μm.

*STM measurement*: To protect the 1*T*-$TaTe_2$ films from exposure to air during the transfer to the STM chamber, we sequentially deposited Te and Se capping layers with the thickness of ~100 nm on the film before taking the samples out of the UHV system of Beamline 10.0.1. After transferal of the sample through the air to the STM UHV chambers, the samples were annealed at 300 ˚C for 2 hours in the UHV system to remove the capping layer before STM measurement. All STM/STS measurements were performed in a commercial Omicron LT-STM held at $T = 4.7$ K. STM tips were prepared on a Cu(111) surface before each set of measurements to avoid tip artifacts. All STM images were edited using WSxM software [48].

*Density functional theory calculations*: All calculations are carried out using density-functional theory (DFT) with the plane-wave-based Vienna *Ab-initio* Simulation Package (VASP) [49]. The projected augmented wave (PAW) method [50] was used to approximate the electron-ion potential. Exchange-correlation effects were treated within the Perdew-Burke-Ernzerhof (PBE) [51] functional form of the generalized gradient approximation (GGA). We used the kinetic energy cut off of 500 eV, and $24 \times 24 \times 1$, $12\times 12 \times 1$, $10\times 10 \times 1$, and $4 \times 4 \times 1$ $\Gamma$-centered $k$-point meshes for the Brillouin zone integration of 1T-$TaX_2$ (X = S, Se, Te), $3 \times 3$, $\sqrt{13} \times \sqrt{13}$ and $\sqrt{19} \times \sqrt{19}$ CDW phases of $1T$-$TaTe_2$, respectively. The calculations were converged in energy to $10^{-6}$ eV/cell, and the structures were relaxed until the forces were less than $10^{-2}$ eV/Å. A large vacuum space of ≥16 Å in the direction of c is applied to avoid any spurious interaction between periodically repeated. The phonon dispersions have been calculated using VASP + Phonopy based on the finite displacement method [52]. To investigate the CDW instability, the electron susceptibilities were calculated. Here the real part of the electron susceptibility is defined as

$$\chi'(q) = \sum_k \frac{f(\varepsilon_k) - f(\varepsilon_{k+q})}{\varepsilon_k - \varepsilon_{k+q}}$$

, where $f(\varepsilon_k)$ is the Fermi-Dirac function. The electron-phonon couplings are obtained using the QUANTUM ESPRESSO package [53], energy cutoff of 476 eV (35 Ry), and a $q$-grid of $12 \times 12 \times 1$.

**Supporting Information**

Supporting Information is available from the Wiley Online Library or from the author.

**Acknowledgments**

The work performed at the Stanford Institute for Materials and Energy Sciences and Stanford University (thin film characterization) was supported by the Office of Basic Energy Sciences, the US Department of Energy under Contract No. DE-AC02-76SF00515. The work performed at the Advanced Light Source (sample growth and ARPES) was supported by the Office of Basic Energy Sciences, the US Department of Energy under Contract No. DE-AC02-05CH11231. The work performed at UC Berkeley (STM measurements) was supported as part of the Center for Novel Pathways to Quantum Coherence in Materials, an Energy Frontier Research Center funded by the US Department of Energy, Office of Science, Basic Energy Sciences. The work at Pusan National University is supported by the National Research Foundation of Korea (NRF) grant funded by the Korea government (MSIP) (Grant No. 2021R1A2C1004266 and 2020K1A3A7A09080369) (photoemission spectroscopy measurements) and Korea Basic Science Institute (National research Facilities and Equipment Center) grant funded by the Ministry of Education (Grant No. 2021R1A6C101A429) (total energy and phonon calculations). K. K was supported by KAERI internal R&D program (524460-22) (first principles calculations). H. R. acknowledges support from the National Research Foundation of Korea (NRF) grant funded by the Korea government (MSIT) (Grant No. 2021R1A2C2014179) (ARPES analysis). C. Z. acknowledges support from a Kavli ENSI Philomathia Graduate Student Fellowship.

Received: ((will be filled in by the editorial staff))
Revised: ((will be filled in by the editorial staff))
Published online: ((will be filled in by the editorial staff))

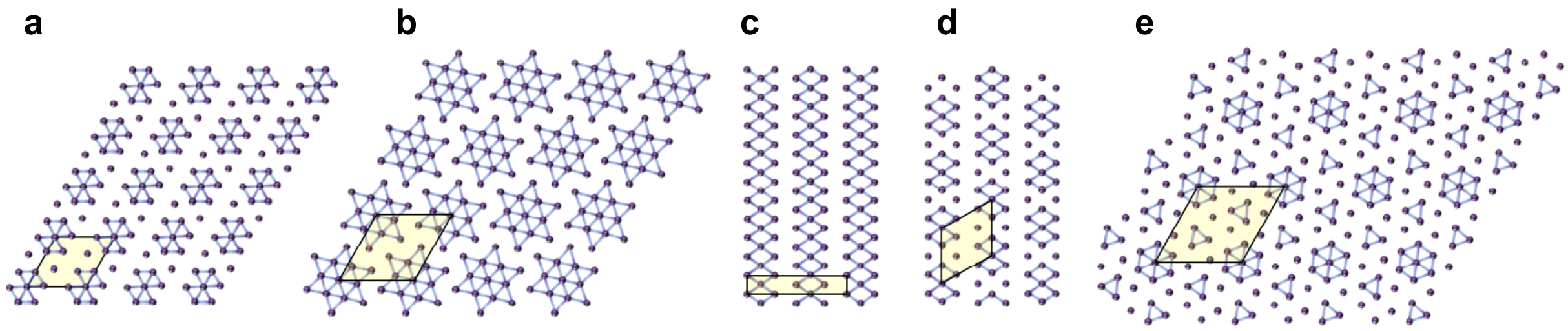

**Figure 1. Possible CDW orders in 1*T*-TaTe$_2$.** Schematic illustrations of the crystal structures with the various possible CDW orders in TMDCs. Purple circles are the transition metals in the middle layer of TMDCs, and the blue lines represent the atomic bonds whose lengths change due to the CDW transition. (a) $3 \times 3$, (b) $\sqrt{13} \times \sqrt{13}$, (c) $3 \times 1 \times 3$ with double zigzag chain, (d) $3 \times 3 \times 3$ with double zigzag chain, and (e) $\sqrt{19} \times \sqrt{19}$ superstructures. Yellow boxes indicate the unit cell in the CDW state.

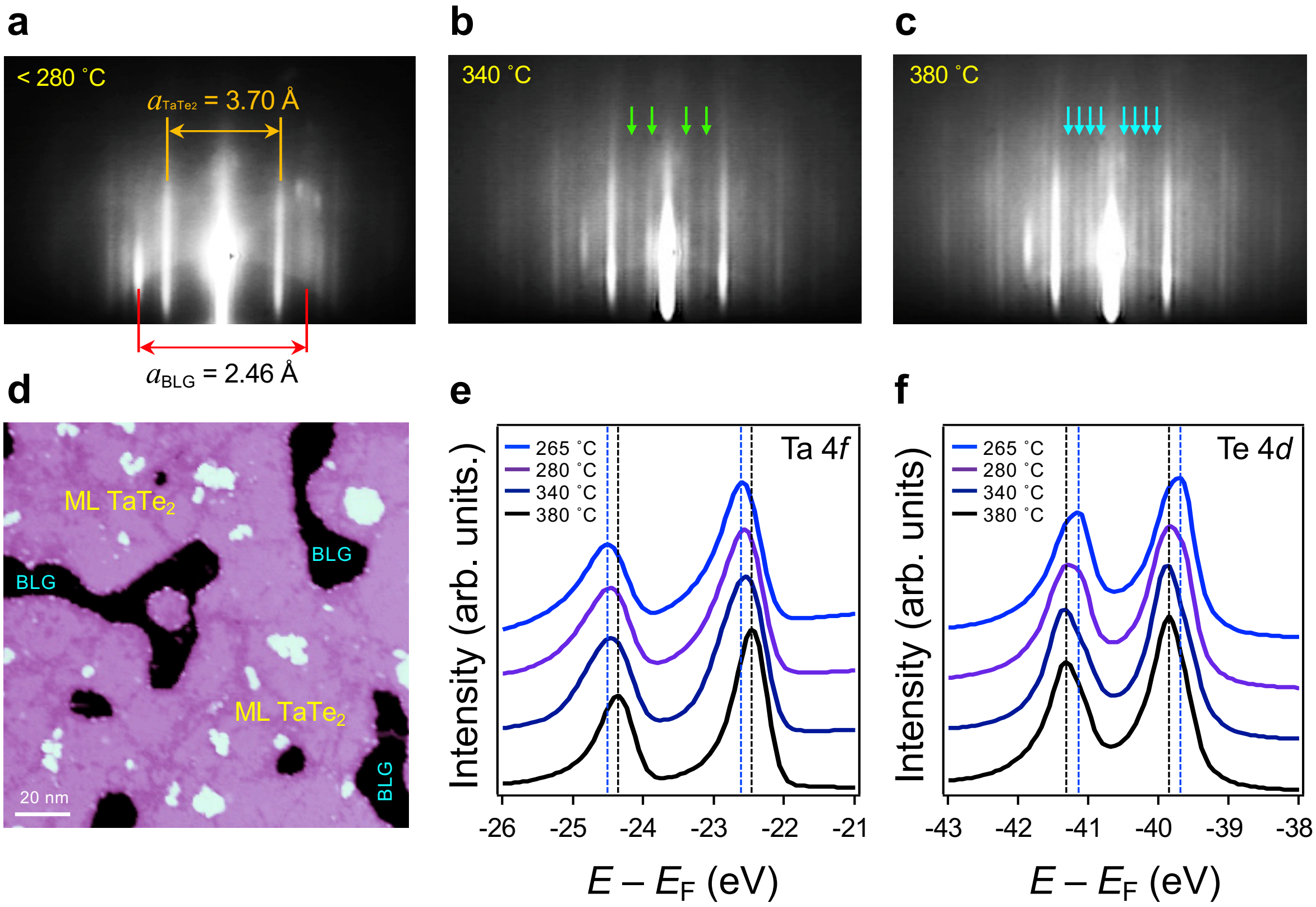

**Figure 2. Characterization of the epitaxially grown ML 1*T*-TaTe$_2$.** (a-c) RHEED images of sub-ML 1*T*-TaTe$_2$ grown on BLG substrate, with post-growth annealing temperatures (a) < 280 ˚C, (b) 340 ˚C, and (c) 380 ˚C. All images are taken at RT. The red and orange lines represent the RHEED lines from BLG substrate and 1*T*-TaTe$_2$ film, respectively. Additional green and cyan arrows indicate the additional lines coming from the formation of superstructures. (d) Typical STM topographic image of ML 1*T*-TaTe$_2$ on BLG substrate ($V_s$ = -1 V, $I_0$ = 0.01 nA, $T$ = 4.7 K). White areas are the remaining Te capping layer. (e-f) Core level photoemission spectra from (e) Ta 4*f* and (f) Te 4*d* levels of ML 1*T*-TaTe$_2$.

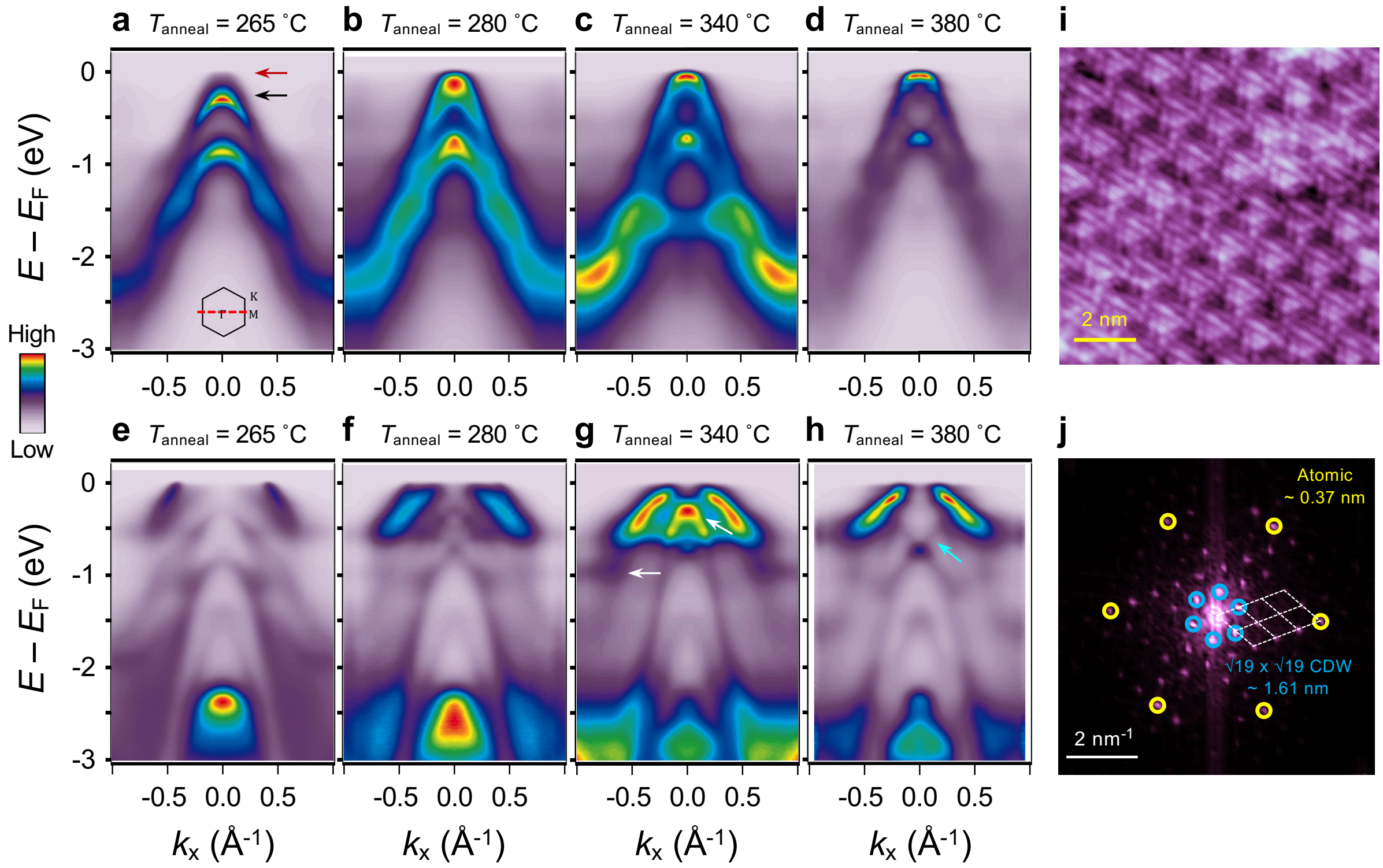

**Figure 3. Polarization- and annealing temperature-dependent ARPES intensity maps and STM images of ML 1*T*-TaTe$_2$.** (a-h) ARPES intensity maps of ML 1*T*-TaTe$_2$ annealed at (a,e) 265 ˚C, (b,f) 280 ˚C, (c,g) 340 ˚C, and (d,h) 380 ˚C taken along the M − Γ − M direction using (a-d) *s*- and (e-h) *p*-polarized photons ($T$ = 13 K). (i) Atomically-resolved STM image and (j) its FT of ML 1*T*-TaTe$_2$ annealed at 380 ˚C ($V_s$ = -0.25 V, $I_0$ = 0.5 nA, $T$ = 4.7 K). Yellow and blue circles represent Bragg and CDW peaks, respectively. The primary reciprocal lattice vectors of ML 1*T*-TaTe$_2$ are defined with $\theta = 120^{\circ}$.

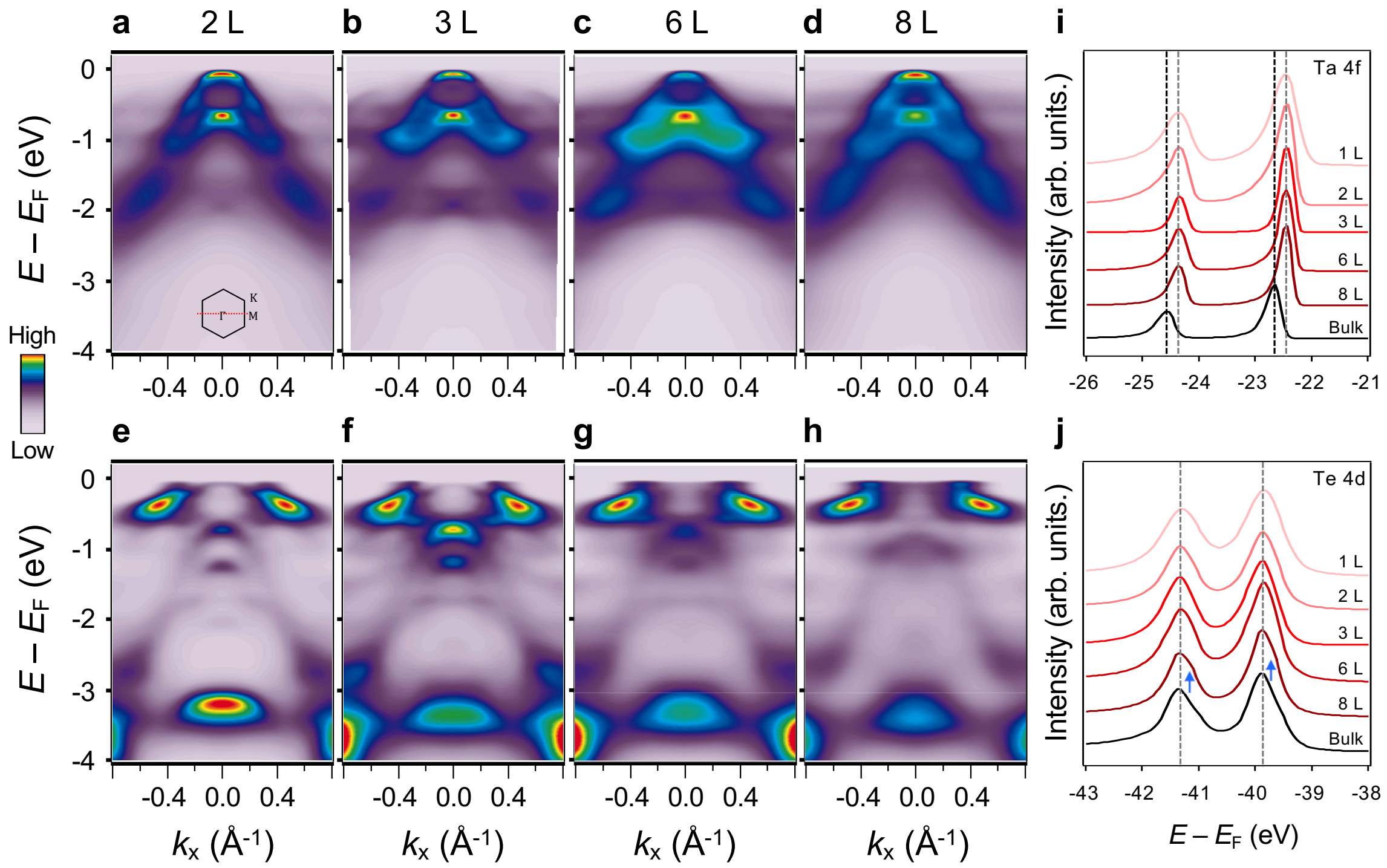

**Figure 4. Thickness-dependent electronic structures of 1*T*-$TaTe_2$ thin film.** (a-h) Polarization-dependent ARPES intensity maps of (a,e) 2 L, (b,f) 3 L, (c,g) 6 L, and (d,h) 8 L 1*T*-$TaTe_2$ taken along the M − Γ − M direction using (a-d) *s*- and (e-h) *p*-polarized photons ($T$ = 13 K). (i,j) Thickness-dependent core level PES spectra from (i) Ta 4*f* and (j) Te 4*d* levels of 1*T*-$TaTe_2$.

**Table I. Relative total energy difference (meV) for the 1*T* monolayer structures with $\mathbf{3 \times 3}$, $\mathbf{\sqrt{13} \times \sqrt{13}}$, and $\mathbf{\sqrt{19} \times \sqrt{19}}$ unit cells.**

| Unit cell | $\mathbf{3 \times 3}$ | $\mathbf{\sqrt{13} \times \sqrt{13}}$ | $\mathbf{\sqrt{19} \times \sqrt{19}}$ |
|---|---|---|---|
| 1*T*-$TaS_2$ | - | -28 | -13 |
| 1*T*-$TaSe_2$ | -29 | -67 | -24 |
| 1*T*-$TaTe_2$ | -106 | -96 | -69 |

$$\Delta E\ (meV) = \frac{E[TaX_2(unit\ cell)]}{formula\ unit} - E[1T\text{-}TaX_2]$$

# Supporting Information

**A novel $\sqrt{19} \times \sqrt{19}$ superstructure in epitaxially grown 1*T*-$TaTe_2$ thin films**

*Jinwoong Hwang†*, Yeongrok Jin†, Canxun Zhang†, Tiancong Zhu†, Kyoo Kim, Yong Zhong, Ji-Eun Lee, Zongqi Shen, Yi Chen, Wei Ruan, Hyejin Ryu, Choongyu Hwang, Jaekwang Lee, Michael F. Crommie, Sung-Kwan Mo* and Zhi-Xun Shen**

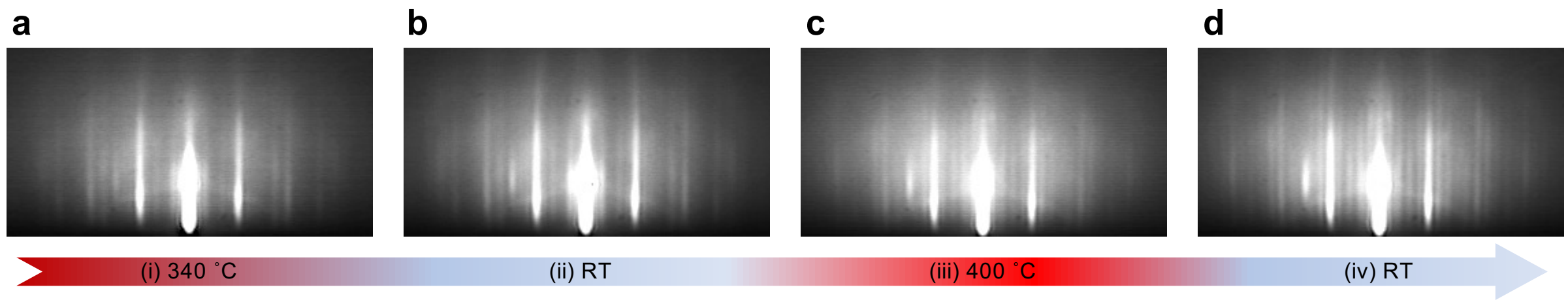

**Figure S1. Transformation of RHEED patterns by annealing ML 1*T*-$TaTe_2$.** (a,b) RHEED images taken (a) at 340 ˚C and (b) at RT after cooling down to RT. (c,d) RHEED images taken (c) at 400 ˚C after annealing again up to 400 ˚C, and (d) at RT after cooling down to RT.

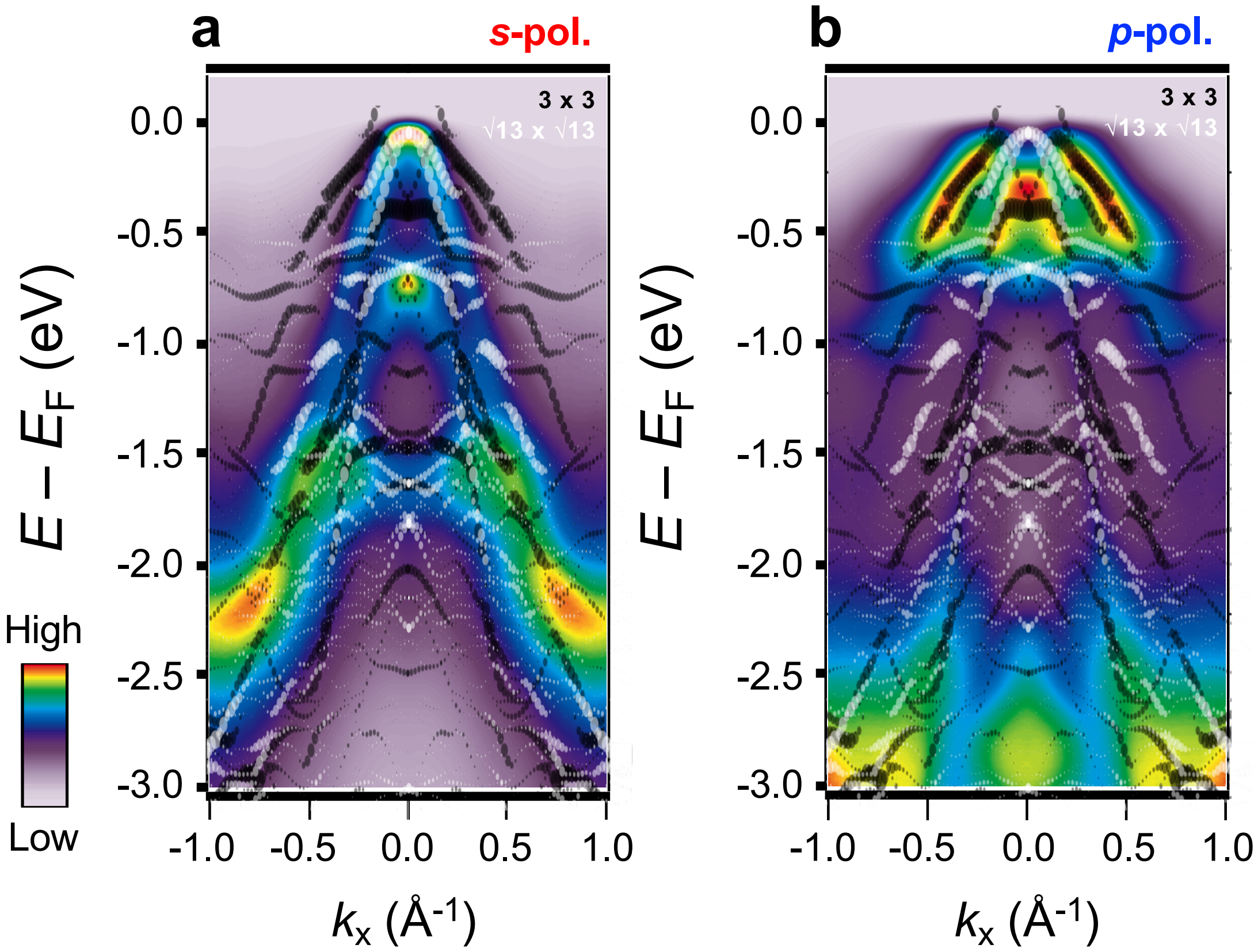

**Figure S2. Coexistence of 3 × 3 and $\sqrt{13} \times \sqrt{13}$ CDW orders at $T_{anneal}$ = 340 ˚C in ML 1*T*-$TaTe_2$.** ARPES intensity maps measured with (a) *s*- and (b) *p*-polarized photons. DFT calculations for 3 × 3 (black) and $\sqrt{13} \times \sqrt{13}$ (white) CDW orders are overlapped.

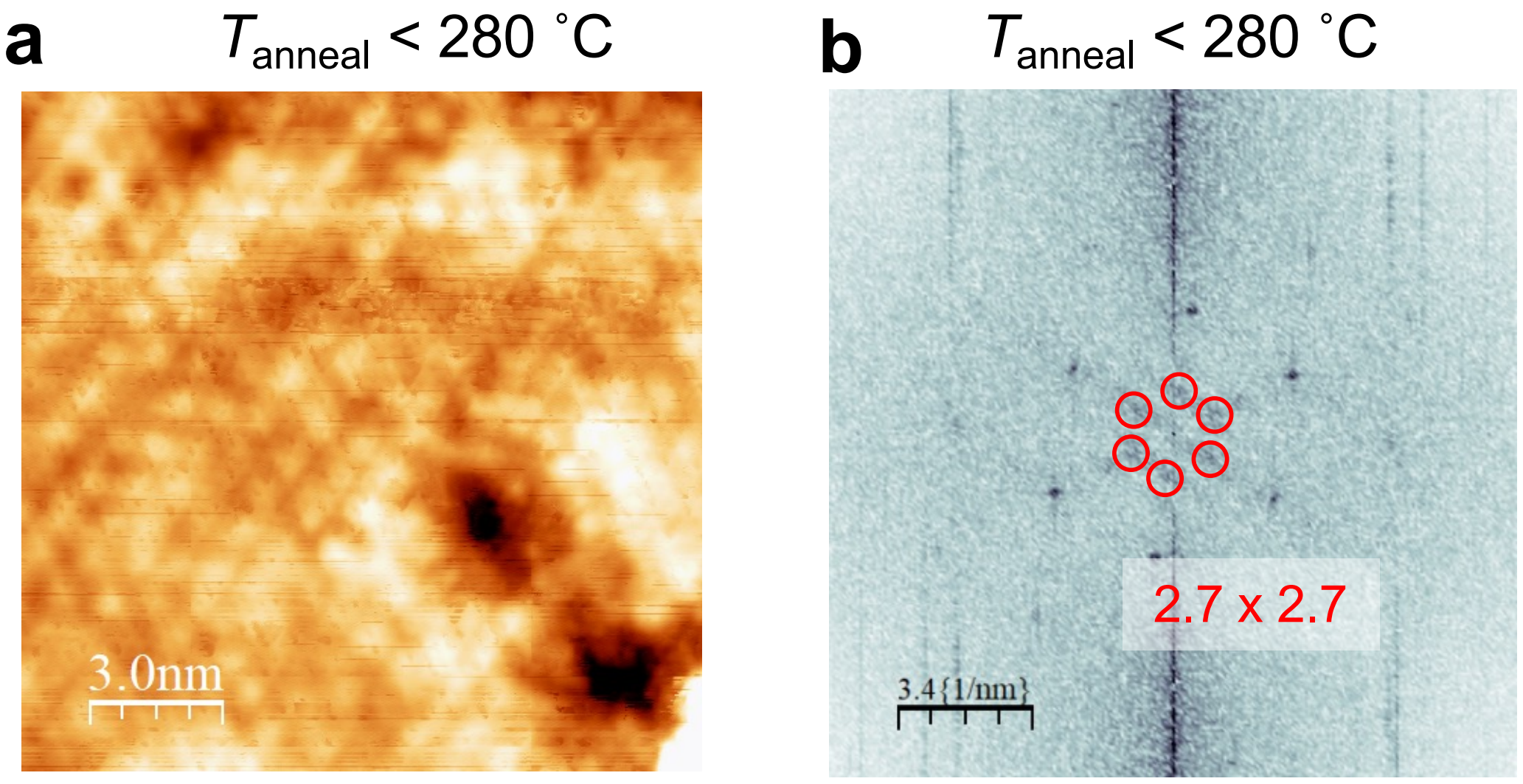

**Figure S3. STM measurements for ML 1*T*-$TaTe_2$ at $T_{anneal}$ < 280 ˚C.** (a) Atomically-resolved STM images of ML 1*T*-$TaTe_2$ annealed below 280 ˚C and (b) its FT ($V_s$ = -0.02 V, $I_0$ = 0.5 nA, $T$ = 4.7 K). Red circles represent 2.7 × 2.7 superstructure peaks.

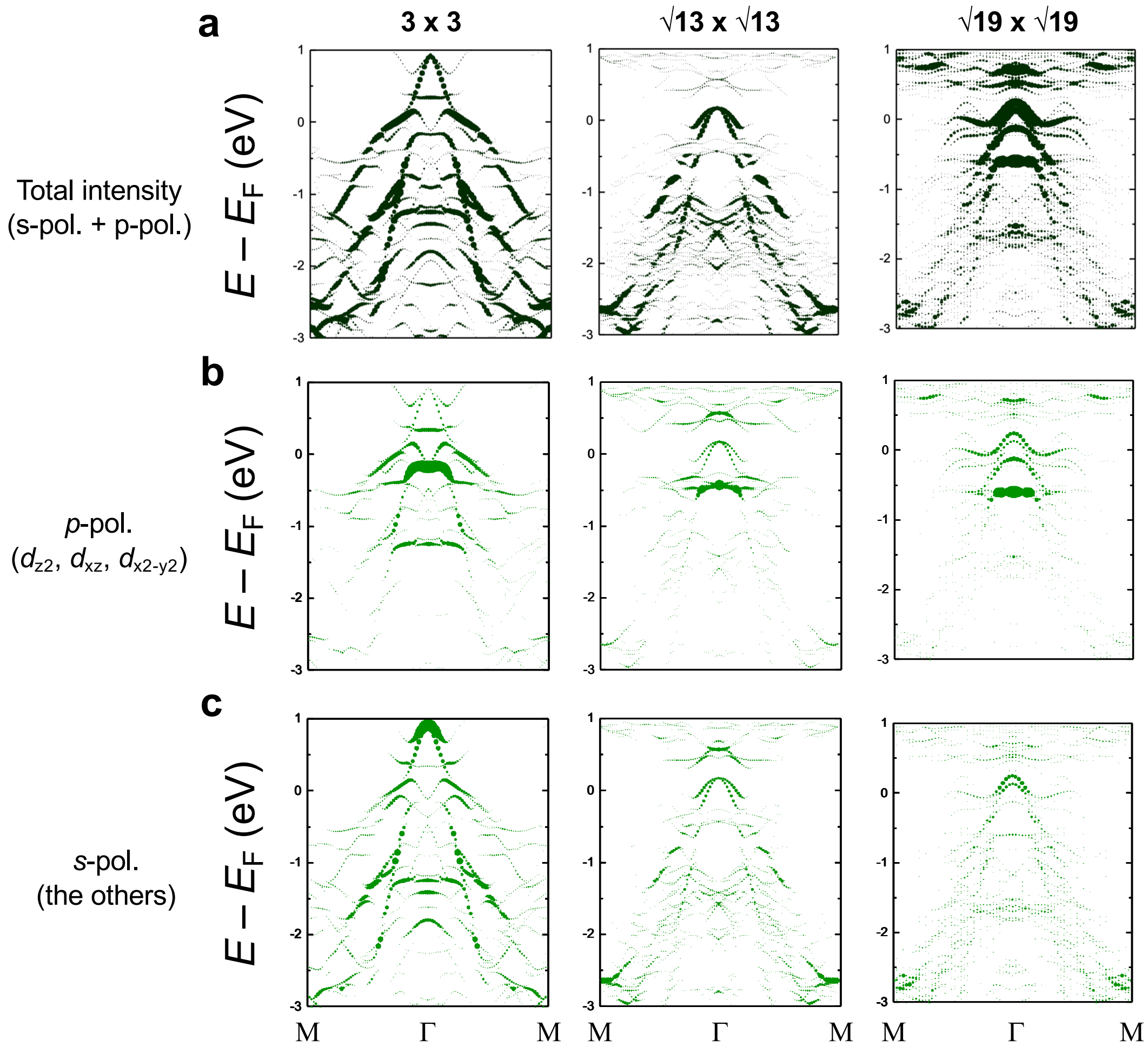

**Figure S4. Orbital-selective DFT band structures for various CDW orders in ML 1*T*-$TaTe_2$** (a) Total, (b) *p*-polarized, and (c) s-polarized photon sensitive orbital-projection DFT band structures. Left: $3 \times 3$, middle: $\sqrt{13} \times \sqrt{13}$, and right: $\sqrt{19} \times \sqrt{19}$ CDW orders for ML 1*T*-$TaTe_2$.

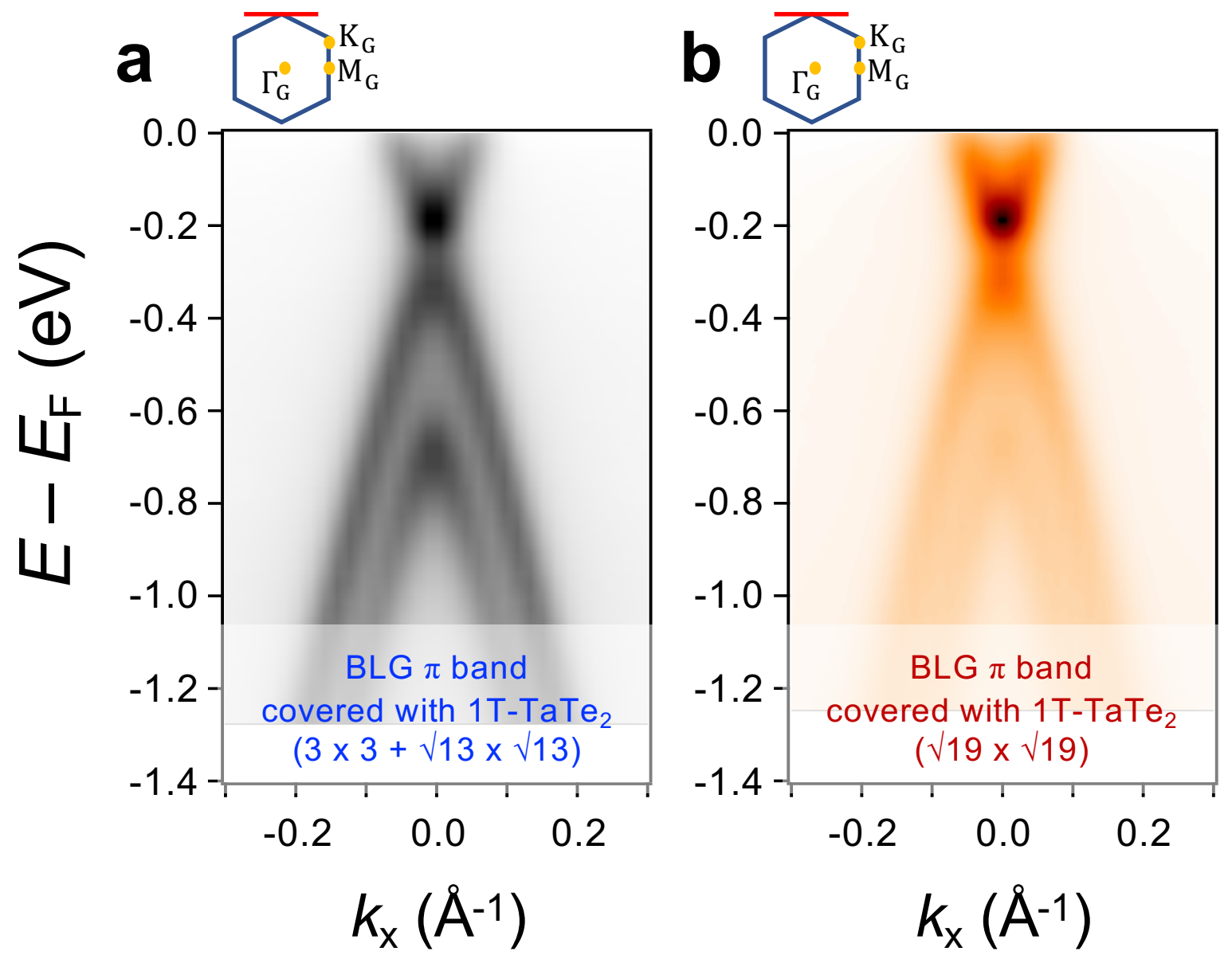

**Figure S5. ARPES spectra of BLG *π* bands near *E*F.** ARPES intensity maps taken at the K point perpendicular to the Γ-K direction covered with ML 1*T*-$TaTe_2$ form with (a) $(3 \times 3 + \sqrt{13} \times \sqrt{13})$ and (b) $\sqrt{19} \times \sqrt{19}$ CDW orders.

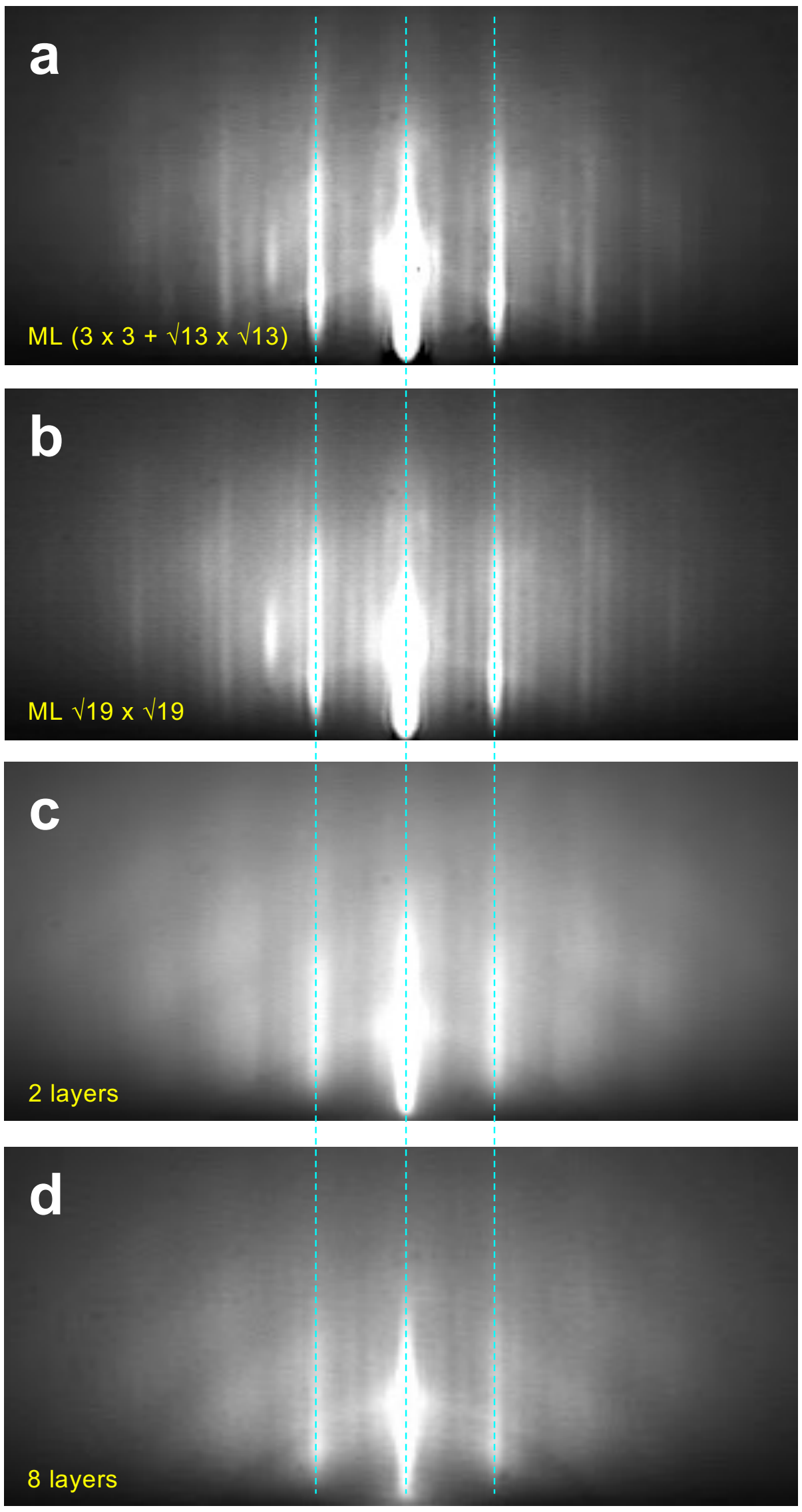

**Figure S6. RHEED images for MBE-grown 1*T*-$TaTe_2$ thin films.** (a,b) RHEED images for ML 1*T*-$TaTe_2$ form with (a) $(3 \times 3 + \sqrt{13} \times \sqrt{13})$ and (b) $\sqrt{19} \times \sqrt{19}$ CDW orders. (c,d) RHEED images for (c) 2 L and (d) 8 L 1*T*-$TaTe_2$, respectively. Cyan-dashed lines are the guides to eyes.

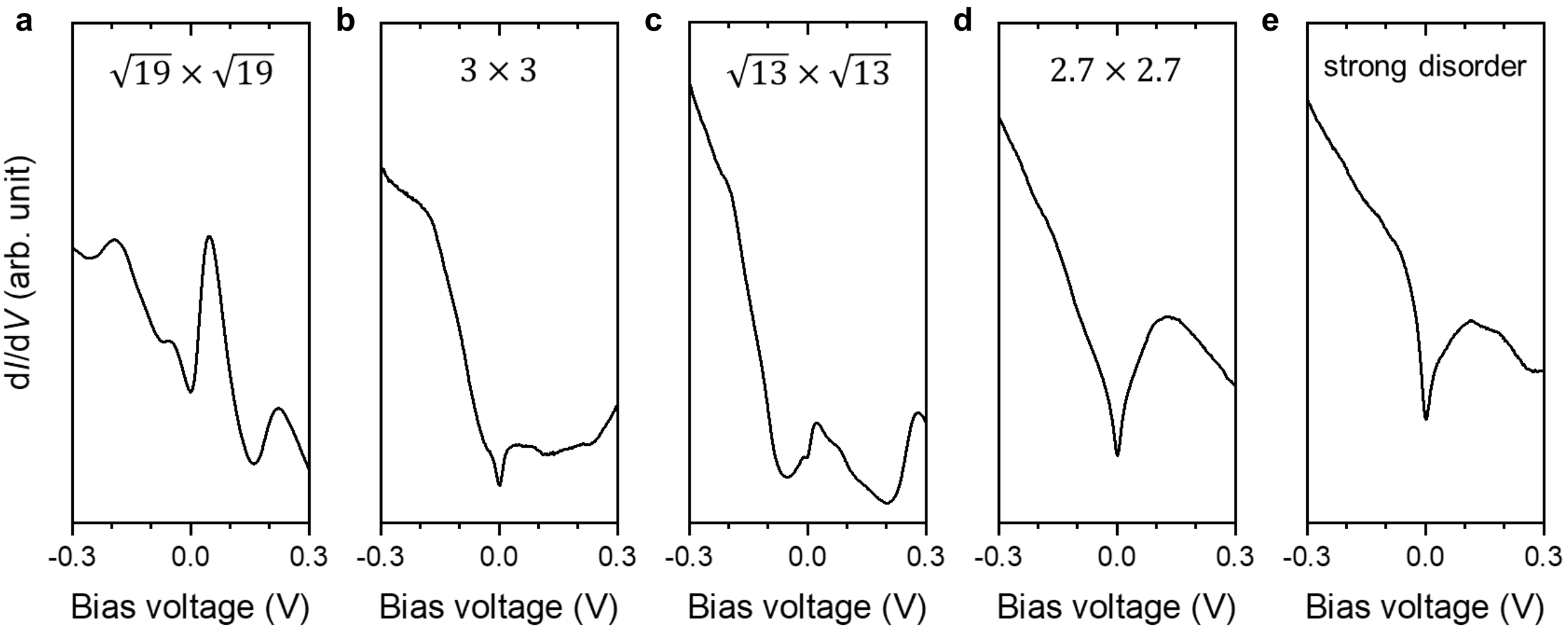

**Figure S7. STM d*I*/d*V* spectra for various phases in ML 1*T*-$TaTe_2$.** STM dI/dV spectra obtained at (a) $\sqrt{19} \times \sqrt{19}$ ($T_{anneal}$ = 380 ˚C), (b) $3 \times 3$, (c) $\sqrt{13} \times \sqrt{13}$ ($T_{anneal}$ = 380 ˚C), (d) $2.7 \times 2.7$ CDW orders, and (e) strong disorder area ($T_{anneal}$ < 280 ˚C), respectively.

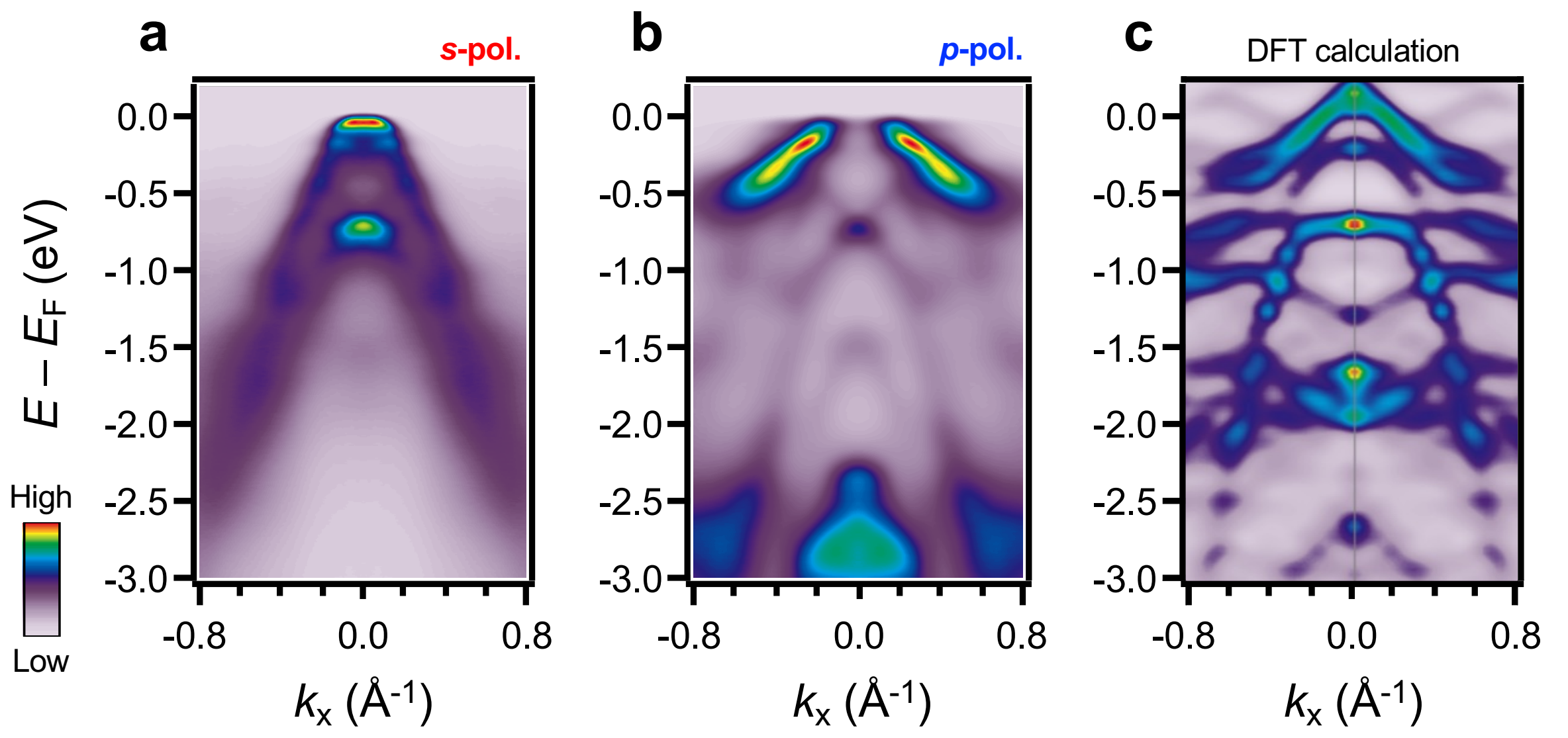

**Figure S8. Comparison between ARPES and DFT band dispersions for $\sqrt{19} \times \sqrt{19}$ CDW order in ML 1*T*-TaTe$_2$.** ARPES intensity maps of ML 1*T*-TaTe$_2$ with $\sqrt{19} \times \sqrt{19}$ CDW order taken with (a) *s*- and (b) *p*-polarized photons. (c) Calculated DFT band structure for $\sqrt{19} \times \sqrt{19}$ CDW order in ML 1*T*-TaTe$_2$.

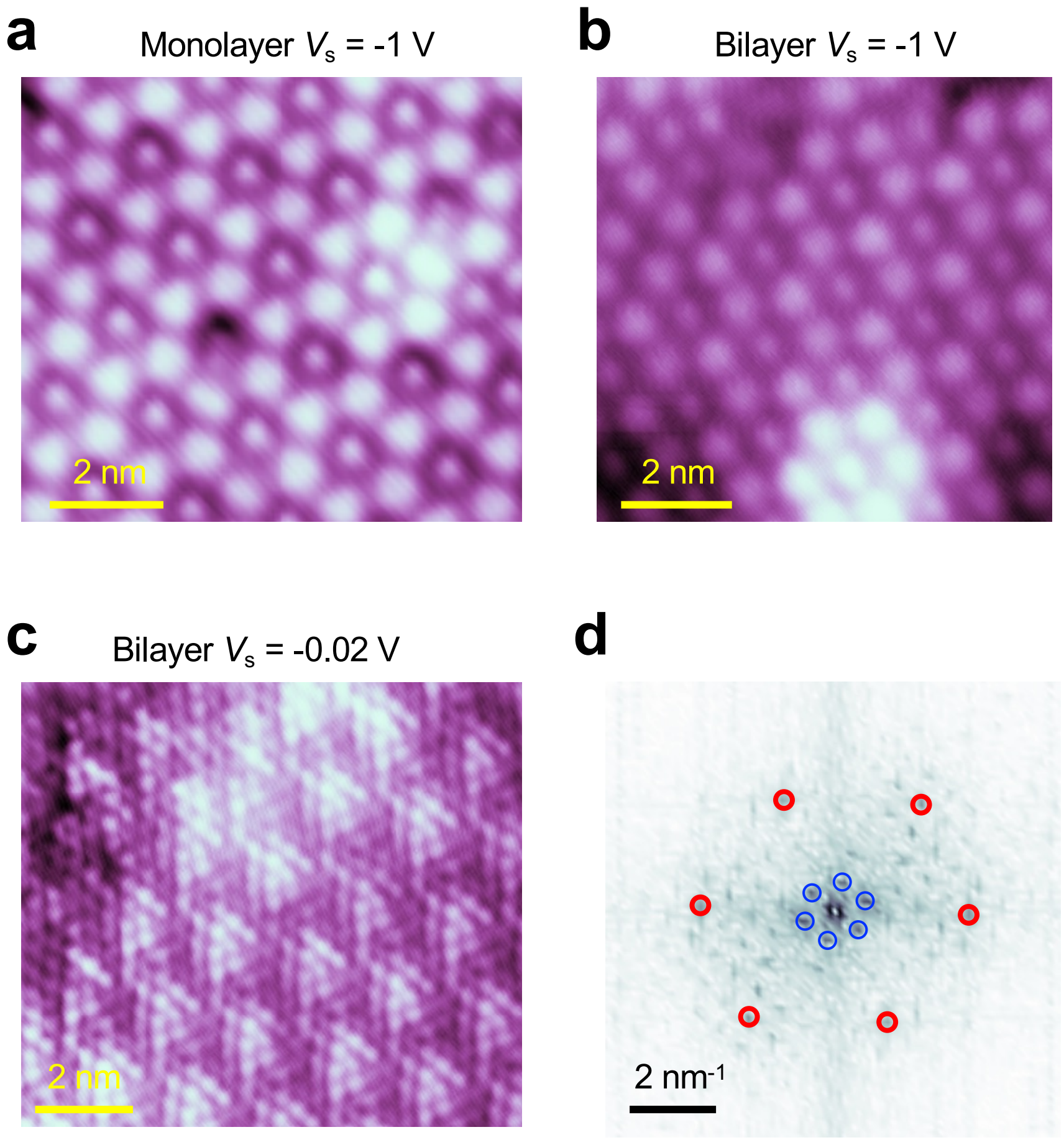

**Figure S9. STM measurements for BL 1*T*-$TaTe_2$.** (a,b) Atomically-resolved STM images of (a) ML and (b) BL 1*T*-$TaTe_2$ annealed at 380 ˚C ($V_s$ = -1 V, $I_0$ = 0.5 nA, $T$ = 4.7 K). (c) Atomically-resolved STM images taken at $V_s$ = -0.02 V and (d) its FT of BL 1*T*-$TaTe_2$ annealed at 380 ˚C ($V_s$ = -0.02 V, $I_0$ = 0.5 nA, $T$ = 4.7 K). Red and blue circles represent Bragg and $\sqrt{19} \times \sqrt{19}$ CDW peaks, respectively.

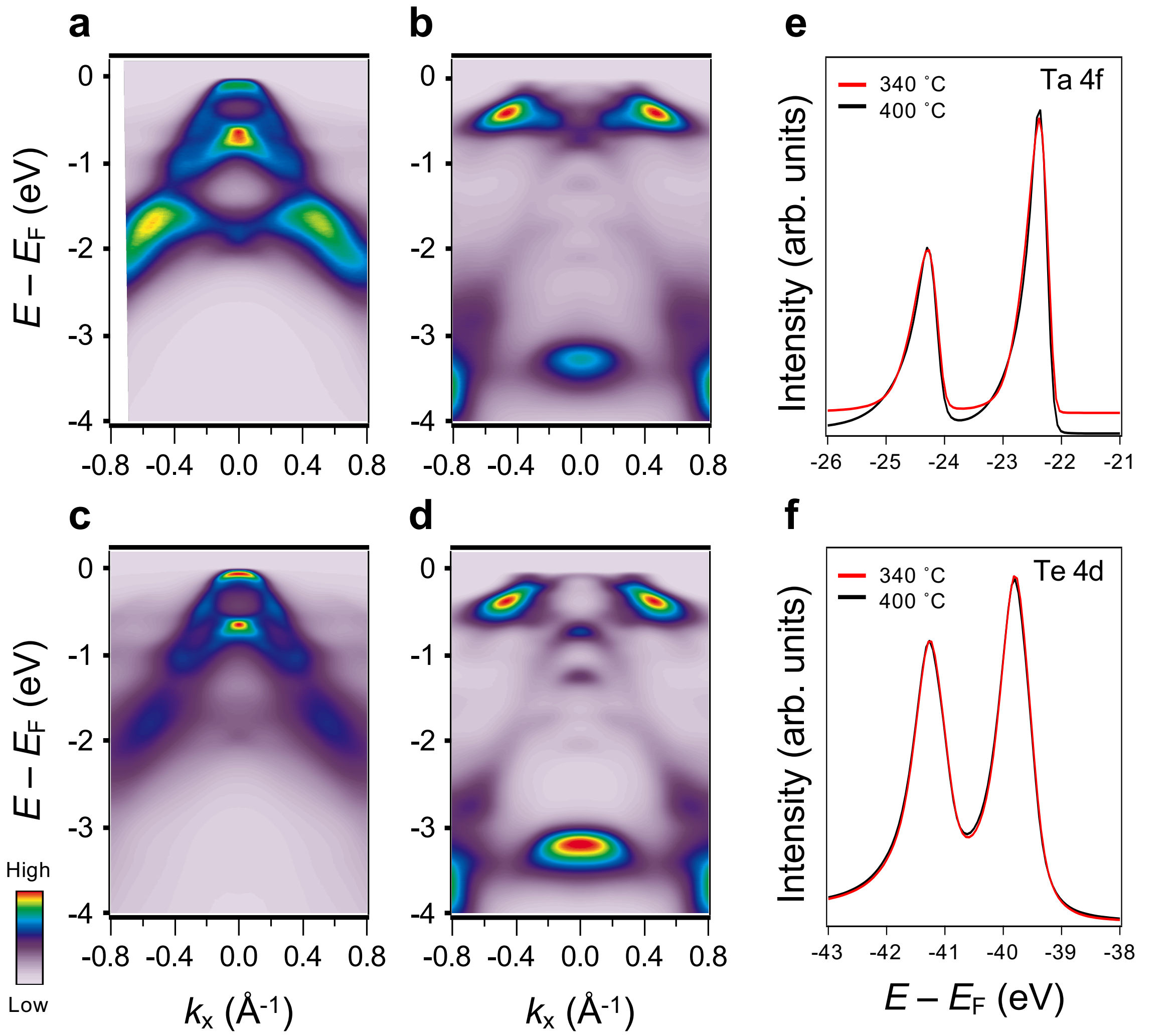

**Figure S10. Annealing temperature-dependent ARPES intensity maps and core level measurements for BL 1*T*-$TaTe_2$.** (a,b) ARPES band structure of BL 1*T*-$TaTe_2$ annealed at 340 ˚C using (a) *s*- and (b) *p*-polarized photons. (c,d) ARPES band structure of BL 1*T*-$TaTe_2$ annealed at 400 ˚C using (c) *s*- and (d) *p*-polarized photons. (e,f) Core level spectra from (e) Ta 4*f* and (f) Te 4*d* of BL 1*T*-$TaTe_2$, respectively.

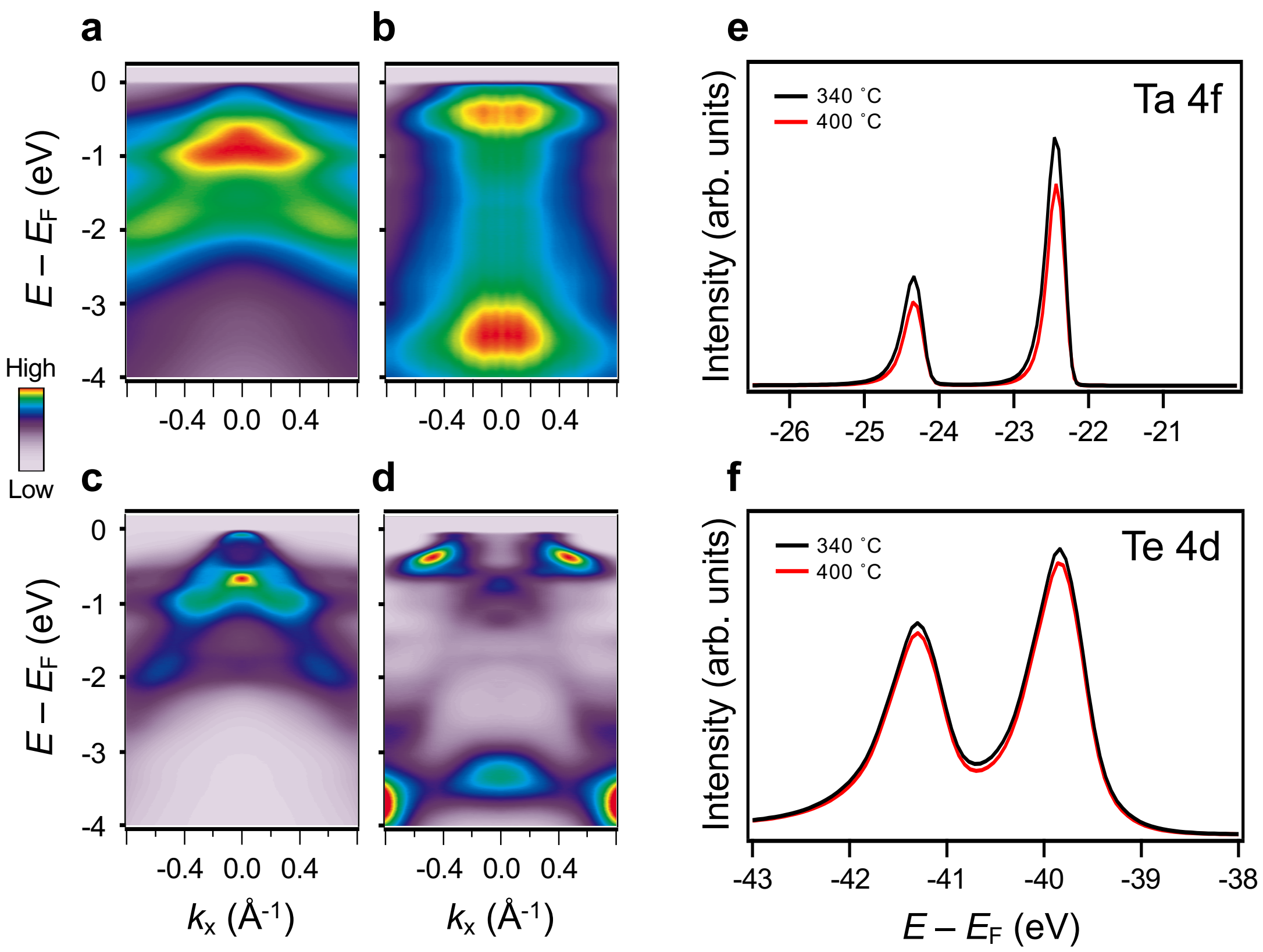

**Figure S11. Annealing temperature-dependent ARPES intensity maps and core level measurements for 6 L 1*T*-$TaTe_2$.** (a,b) ARPES band structure of 6 L 1*T*-$TaTe_2$ annealed at 340 ˚C using (a) *s*- and (b) *p*-polarized photons. (c,d) ARPES band structure of BL 1*T*-$TaTe_2$ annealed at 400 ˚C using (c) *s*- and (d) *p*-polarized photons. (e,f) Core level spectra from (e) Ta 4*f* and (f) Te 4*d* of BL 1*T*-$TaTe_2$, respectively.

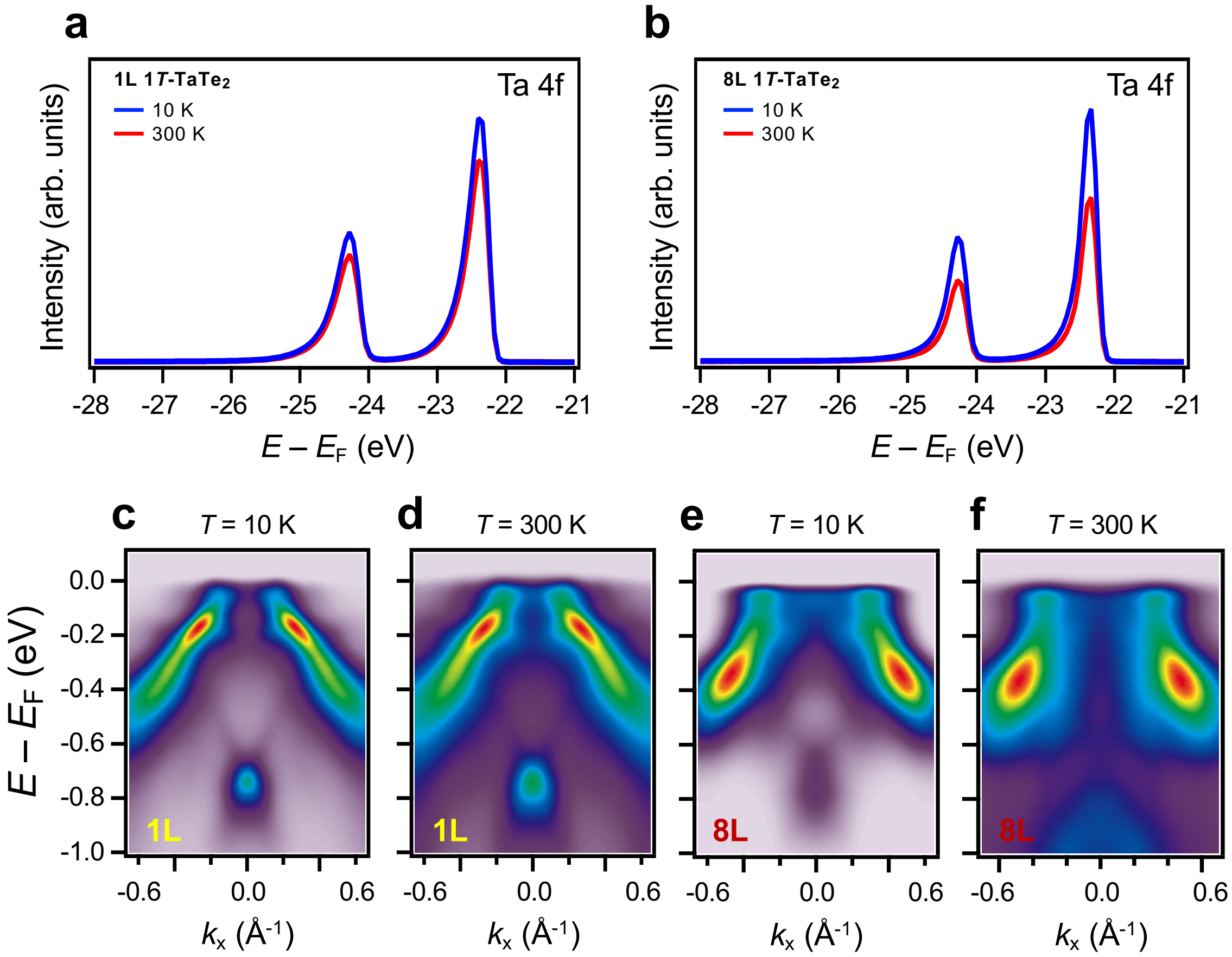

**Figure S12. Temperature-dependent ARPES intensity maps and core level measurements for ML and 8 L 1*T*-TaTe₂.** (a,b) Ta 4*f* core level for (a) ML and (b) 8 L 1*T*-TaTe₂ taken at 10 K (blue) and 300 K (red). (c-f) ARPES band structure of (c,d) ML and (e,f) 8 L 1*T*-TaTe₂ taken at (c,e) 10 K and (d,f) 300 K using *p*-polarized photons, respectively.

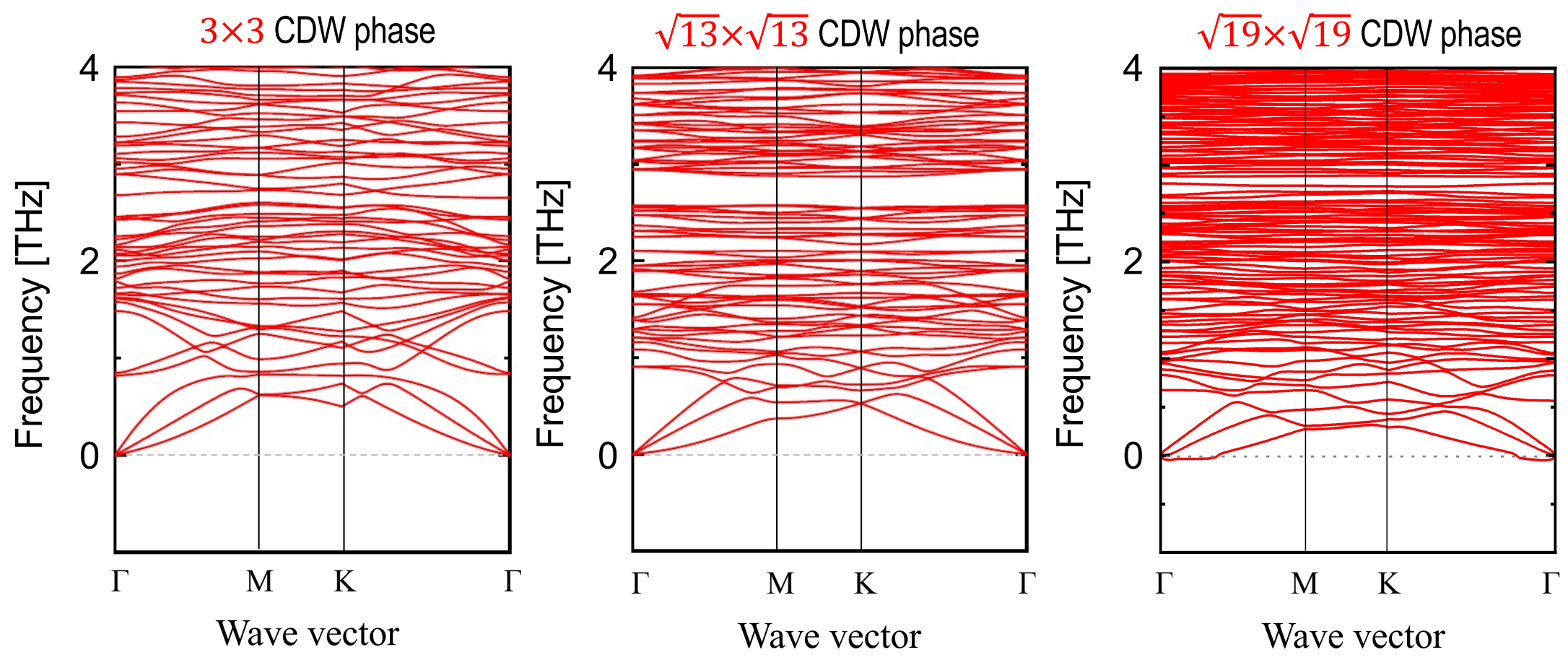

**Figure S13. Phonon dispersions of $3 \times 3$, $\sqrt{13} \times \sqrt{13}$, and $\sqrt{19} \times \sqrt{19}$ CDW phases in ML 1*T*-$TaTe_2$.**

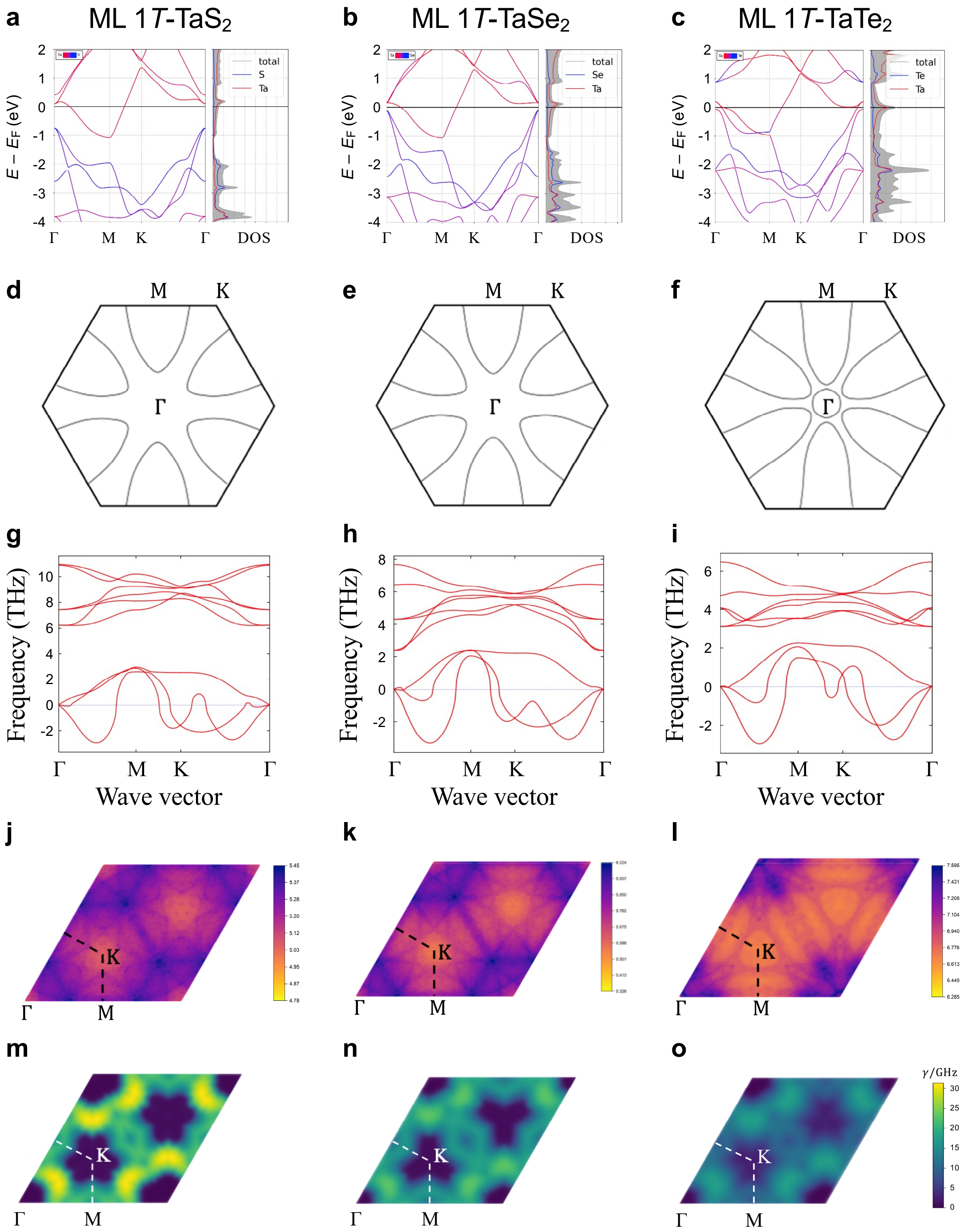

**Figure S14. The electronic structures, phonon spectra and electron susceptibilities of ML 1*T*-TaCh₂ (Ch = S, Se, Te).** (a-c) Electronic band structures and density of states of the ML 1*T*-$TaCh_2$ for S, Se, Te. The red and blue lines are the contributions from the Ta and the chalcogen atoms. The contribution at the Fermi energy is mostly from the Ta atom. (d-f) Fermi surface, (g-i) phonon dispersions, (j-l) the real part of the electron susceptibility, (m-o) contour map of the phonon linewidth $\gamma$ of the lowest phonon modes of the ML 1*T*-$TaCh_2$.

**Electron and phonon properties of ML 1*T*-$TaCh_2$ (Ch = S, Se, Te)**

We investigated the electronic structures and phonon dispersions of ML 1*T*-$TaCh_2$ (Ch = S, Se, and Te) to reveal the driving force of the obtained CDW orders. Our calculations show that ML 1*T*-$TaS_2$ and 1*T*-$TaSe_2$ have almost the same shape of the density of states (DOS) (**Figures S14a and S14b**) and Fermi surfaces (**Figures S14d and S14e**), while ML 1*T*-$TaTe_2$ exhibits a sharp peak of DOS right at $E_F$ (**Figure S14c**) and a hole pocket in the Fermi surface (**Figure S14f**). As can be seen near Γ point in band structures, this is due to the enhanced mixing of *p*-*d* orbitals by the extended Te 5*p* orbital. Despite the different electronic structures of 1*T*-$TaCh_2$, however, phonon dispersions are quite similar, where the presence of phonons with imaginary frequency at Γ–K and Γ–M directions indicates that the lattice is unstable.

We have further calculated electron susceptibility and phonon linewidth, which are conventionally investigated to reveal the driving force of the CDW transitions in TMDCs. The electronic susceptibility (**Figures S14j-S14l**) shows a dominant peak at $0.22\overline{\Gamma M}$ for ML 1*T*-$TaS_2$, $0.26\overline{\Gamma M}$ for 1*T*-$TaSe_2$, and $0.23\overline{\Gamma M}$ for 1*T*-$TaTe_2$, respectively. These values deviate from not only the experimentally obtained $\sqrt{13} \times \sqrt{13}$ CDW order in ML 1*T*-$TaCh_2$, but also $3 \times 3$ and $\sqrt{19} \times \sqrt{19}$ CDW orders in ML 1*T*-$TaTe_2$. On the other hand, all phonon spectra show a peak at $\frac{1}{\sqrt{13}}\ \overline{\Gamma K}$ (**Figures S14m-S14o**). This result is consistent with the bulk 1*T*-$TaS_2$ and 1*T*-$TaSe_2$, indicating that $\sqrt{13} \times \sqrt{13}$ CDW order in ML 1*T*-$TaCh_2$ is driven by strong electron-phonon coupling rather than Fermi surface nesting. However, both electronic susceptibility and phonon spectra do not directly show any connection to $3 \times 3$ and $\sqrt{19} \times \sqrt{19}$ CDW orders. Instead, the phonon spectra of ML 1*T*-$TaTe_2$ (**Figure S14o**) are much broader compared to 1*T*-$TaS_2$ and 1*T*-$TaSe_2$. Since the broad linewidth peak of ML 1*T*-$TaTe_2$ includes the ***q***-points of $3 \times 3$ and $\sqrt{19} \times \sqrt{19}$ CDW orders, we can not fully exclude the strong electron-phonon coupling as a main driving force.